# Unveiling the hidden nematicity and spin subsystem in FeSe

Chih-Wei Luo[1,2*], Po Chung Cheng[1], Shun-Hung Wang[3], Jen-Che Chiang[3], Jiunn-Yuan Lin[3*], Kaung-Hsiung Wu[1], Jenh-Yih Juang[1], Dmitry A. Chareev[4,6], Olga S. Volkova[5,6,7] and Alexander N. Vasiliev[5,6,7]

*1. Department of Electrophysics, National Chiao Tung University, Hsinchu 300, Taiwan*

*2. Taiwan Consortium of Emergent Crystalline Materials, Ministry of Science and Technology, Taipei 10601, Taiwan*

*3. Institute of Physics, National Chiao Tung University, Hsinchu 300, Taiwan*

*4. Institute of Experimental Mineralogy, Russian Academy of Sciences, 142432, Chernogolovka, Moscow District, Russia*

*5. Physics Faculty, Moscow State University, Moscow 119991, Russia*

*6. Institute of Physics and Technology, Ural Federal University, Mira st. 19, 620002 Ekaterinburg, Russia*

*7. National University of Science and Technology "MISiS", Moscow 119049, Russia*

*Correspondence: C W Luo (cwluo@mail.nctu.edu.tw) or J-Y Lin (ago@cc.nctu.edu.tw)

**The nematic order (nematicity) is considered one of the essential ingredients to understand the mechanism of Fe-based superconductivity. In most Fe-based superconductors (pnictides), nematic order is reasonably close to the antiferromagnetic order. In FeSe, in contrast, a nematic order emerges below the structure phase transition at $T_s$ = 90 K with no magnetic order. The case of FeSe is of paramount importance to a universal picture of Fe-based superconductors. The polarized ultrafast spectroscopy provides a tool to probe simultaneously the electronic structure and the magnetic interactions through quasiparticle dynamics. Here we show that this approach reveals both the electronic and magnetic**

**nematicity below and, surprisingly, its fluctuations far above $T_s$ to at least 200 K. The quantitative pump-probe data clearly identify a correlation between the topology of the Fermi surface (FS) and the magnetism in all temperature regimes, thus providing profound insight into the driving factors of nematicity in FeSe and the origin of its uniqueness.**

**INTRODUCTION**

The progress in understanding Fe-based superconductors has formed a most intriguing chapter in modern condensed matter physics[1-3]. The existence of nematic order has become well established in Fe-based superconductors and is considered an essential ingredient to understand the mechanism of Fe-based superconductivity [4-6]. The nematic order breaks the rotational symmetry by making the $x$ and $y$ directions in the plane non-equivalent, while preserving the time-reversal symmetry. The chalcogenide FeSe has a superconducting transition temperature $T_c \sim 8.5$ K; its tetragonal structure undergoes a transition to orthorhombic below $T_s = 90$ K. No long-range magnetic order has ever been detected in FeSe down to the lowest temperatures [7-10]. In this respect, as the structurally simplest Fe-based superconductor, FeSe has unexpectedly emerged in the frontier of Fe-based superconductivity research [11-25]. Up to date, few consensuses have been reached on either the electronic structure of FeSe or its nematic and superconducting mechanisms. For example, very recent angle-resolved photoemission spectroscopy (ARPES) has indicated a small Fermi surface (FS) above $T_s$ that cannot be reproduced by density functional theory calculations [10,14-16,26-28]. Furthermore, various ARPES groups concur on an even smaller FS below $T_s$ [10,14-16,26-30], which is in general consistent with the Sommerfeld coefficient observed from the specific heat [31,32]. Nevertheless, how and why a FS is reconstructed in FeSe through $T_s$ is poorly characterized. As for the nematic order under $T_s$, almost everyone agrees on an electronic origin, as a 0.2% orthorhombic distortion is unlikely to lead to the observed FS elongation and the shift of band energy at the M point [15]. Nevertheless, whether the nematicity in FeSe is magnetically or orbitally driven is under current fierce debate, whereas it is generally considered to be driven by magnetism in pnictides [6]. This controversy occurs largely due to the absence of the magnetic order in FeSe that remains an unsolved puzzle. The existence of nematic fluctuations above $T_s$ is, likewise, not entirely clear in the literature.

In the present work, we utilized the polarized femtosecond pump-probe spectroscopy of FeSe to elucidate the above issues. This probe is relevant to both the charge and spin channels, and is sensitive to fluctuations or the short-range order. For example, the wavelength-dependent femtosecond spectroscopy clearly revealed the magnetic fluctuations at $T$ = 170 K in HoMnO$_3$, far above the long-range antiferromagnetic $T_N$ = 76 K [33]. A similar technique has been applied to detect the nematic fluctuations above $T_s$ in pnictides [34-36]. Here we employed polarized ultrafast spectroscopy to elucidate the detailed orientation and temperature dependence of the quasiparticle dynamics in FeSe. As a results of this comprehensive survey, the hidden nematic fluctuations and spin subsystem in FeSe is unveiled.

**RESULTS**

Figure 1 shows the typical polarization-dependent photoinduced reflectivity ($\Delta R/R$) transients on the (001) plane of an FeSe single crystal at various temperatures. At $T$ = 60 K, below $T_s$, the $\Delta R/R$ transients demonstrate clear nematicity (Fig. 1a) in this phase, as also indicated in other experiments [10,14-16,26-28]. We show below that this nematicity in dynamics reveals information of both the quasiparticle and magnetic channels. Astonishingly, $\Delta R/R$ shows profound nematic fluctuations even at $T$ = 150 K, far above $T_s$ (Fig. 1b); this two-fold symmetry persists up to at least 200 K (Fig. 2c). Overall, the raw data in Fig. 1 indicate clear nematic signals in ultrafast dynamics at the highest temperatures unprecedented in preceding reports. Furthermore, the two-fold symmetry pattern shifts by 90° when the temperature passes through $T_s$ as shown in Fig. 1. To depict the context of Fig. 1 more clearly, Fig. 2 shows the typical $\Delta R/R$ transients with the electric field **E** along $\phi = 0°$ and $\phi = 90°$ at the temperatures associated with those in Fig. 1. (Angles $\phi = 0°$ and $\phi = 90°$ were chosen to represent the largest nematic signals, which are corresponding to *a*-axis and *b*-axis of orthorhombic structure, respectively.) Both the sign and the amplitude of $\Delta R/R$ transients show clear nematicity between $\phi = 0°$ and $\phi = 90°$ at 60 K, as shown in Fig. 2a. With $T$ increasing to 150 K, the sign of $\Delta R/R$ transients along $\phi = 90°$ dramatically reverses from negative to positive. Although this pattern shift was unexpected, it manifests a valuable clue to the coupling between magnetism and the FS topology in FeSe, as discussed below.

The relaxation processes ($t > 0$) of $\Delta R/R$ transients in FeSe single crystals are described phenomenologically with

$$\frac{\Delta R}{R} = A_1 e^{-\frac{t}{\tau_1}} + A_2 e^{-\frac{t}{\tau_2}} + A_0 \tag{1}$$

The first term in the right side of Eq. (1) is the decay of the photoexcited electrons (or quasiparticles, QPs) with an initial population number $A_1$, through phonon coupling with a relaxation time $\tau_1$. The second term pertains to the decay of QPs with an initial population number $A_2$, through spin coupling with a corresponding decay time $\tau_2$. The third term describes the energy loss from the hot spot to the ambient environment on a time scale of microsecond, which is much longer than the period of the measurement (~50 ps) and is hence taken as a constant. The ascriptions of the first and the second terms are due mainly to the time and energy scales of $\tau_1$ and $\tau_2$. (See the sections S2 of

S4 Supplementary Information.)

To depict better the temperature dependence of nematic ultrafast dynamics, we undertook another thorough run of $\Delta R/R$ transient measurements with **E** along both $\phi = 0°$ and $\phi = 90°$. According to Eq. (1), each component was extracted from 290 K to 30 K, as shown in Fig. 3a-d. We discuss first the results for $T \leq T_s$; this nematic phase of FeSe has been defined better than the state of $T > T_s$. For the fast component in $\Delta R/R$, a remarkable difference in the amplitude $A_1$ was observed between $\phi = 0°$ and $\phi = 90°$ in the low-temperature regime, shown in Fig. 3a. For $\phi = 0°$, the sign of $A_{1,0}$ below $T_s$ is positive; in contrast, that of $A_{1,90}$ below $T_s$ is negative for $\phi = 90°$. In the literature, this difference is known to manifest the nematicity of the electronic structure. For example, the anisotropic single-particle and collective excitations in the quasi-1D charge-density wave semiconductor $K_{0.3}MoO_3$ [37] and the *d*-wave symmetry of the superconducting gap in cuprate superconductors YBCO [38-41] have been unambiguously revealed by polarized pump-probe spectroscopy. As intriguingly, the orientation anisotropy is shown also in $\tau_1$ (Fig. 3c). $\tau_{1,90}$ for $\phi = 90°$ (red solid circles) shows a notable divergence near $T_s$; this divergence in the rate of QP relaxation indicates a gap opening, at least on some part of the FS. The presence of a gap in the QP density of states gives rise to a bottleneck for carrier relaxation. The mechanism of the bottleneck is described by the Rothwarf-Taylor model [42]; indeed, the temperature dependence of $\tau_{1,90}$ was perfectly fitted according to that model as denoted by the blue solid line in Fig. 3c. Within the same context, $A_{1,90}$ was also fitted, as shown by the blue solid line in Fig. 3e. Assuming a mean-field-like temperature-dependent $\Delta(T) = \Delta(0) [1-(T/T_s)]^x$, the fit of $A_{1,90}$ leads to a gap amplitude $2\Delta(0) = 8.14k_BT_s = 56$ meV, consistent with the energy splitting between $d_{yz}$ and $d_{xz}$ near the M point in the Brillouin zone revealed from ARPES [27,28]; details of the fitting and discussions are available in section S3 of Supplementary Information. Furthermore, the temperature dependence of $\Delta(T)$ is totally consistent with that of the splitting energy at the M point as shown by the solid stars in Fig. 3e. However, there is no such signature of divergence for $\tau_{1,0}$ near $T_s$, which implies a major difference in carrier dynamics and in the band structure along various **k** orientations in the electronic structure. This discrepancy between $\tau_{1,0}$ and $\tau_{1,90}$ seems puzzling; but it actually fits well into the fascinating ARPES observation that, for $T < T_s$ at M point, a gap is opened along $k_y$, whereas there is no gap opening along $k_x$ (see the illustration of the band structure in Fig. 3f) [27]. It is therefore plausible to assign the directions of 0° and 90° as $x$ and $y$, respectively. The abrupt decrease in $\tau_{1,0}$ at 90 K, i.e., the relaxation of QPs becoming efficient, probably indicates that an increased density of states is involved in the relaxation processes along $k_x$ [43,44]. In this scenario, the results of Figs. 3a and 3c also imply that the reconstruction of FS at the M point occurs mainly near 90 K, with no significant fluctuation of electronic nematicity at the M point above 100 K. As ultrafast spectroscopy is a bulk probe, the present results provide bulk evidence to support the electronic structure according to the surface-sensitive ARPES.

We turn to the slow component in Eq. (1) associated with $A_2$ and $\tau_2$ below $T_s$. The high-energy QPs accumulate in the $d$ conduction band of Fe and release their energy through the emission of longitudinal-optical (LO) phonons within a couple of picoseconds. The QPs (or LO phonons) would subsequently also transfer their energy to the spin subsystem and then disturb the spin ordering on the timescale of tens of picoseconds [45]. This spin-related mechanism was clearly observed in this work as represented by $A_2$ and $\tau_2$ in Figs. 3b and 3d. As the temperature decreases, the stronger interaction between spins further results in an extended $\tau_2$ to disturb the spin subsystem. The nematicity of the slow component is even more pronounced than that of the fast component. While $A_{2,0}$ can be clearly found below $T_s$ and increases with decreasing $T$, there is *no* slow component of $\Delta R/R$ along $\phi = 90°$ (as shown by both the data and the fits in the inset of Fig. 3d). This fact implies that, although there is no magnetic order observed in FeSe down to the lowest temperatures, there does exist a strongly anisotropic spin subsystem with the energy scale of ~72 meV (details are available in section S4 of Supplementary Information) below $T_s$, which is consistent with the value obtained from inelastic neutron scattering [46]. In general, the responsive spin orientation is parallel to the polarization of the pump and probe beams. Most spins hence tend to align along direction $x(a)$ as in most pnictides (see, e.g. References 47-49), albeit with a short-range order. This mechanism opens an additional relaxation channel for QP decay along $\phi = 0°$. About at $T_s$, the divergence in $\tau_2$ implies the setting in of the nematic coupling to the spin subsystem (see Fig. 3d), which is caused by the sudden alignment of spins along $x$ [8] in the nematic phase [9,17,20]. Very recently, NMR experiments have observed the strong low-energy magnetic fluctuations below $T_s$ [47]; neutron-scattering experiments have identified the ($\pi$, 0) fluctuation wave vector [8,9]. The origin of magnetism below $T_s$ in FeSe is likely associated with an imbalance of the occupied electron numbers $n_{xz} > n_{yz}$, which is mainly due to the band splitting at the M point below $T_s$ [50]. Within this context, there exists a coupling between the direction of spins and the orientation of the FS distortion through the nematicity of $n_{xz} - n_{yz}$. The direction of spins would likely follow the elongation direction of FS, as in the case of pnictides [49]. Overall, the results from the present work on the nematic ultrafast dynamics are illustrated in the green-colored nematic order phase of Fig. 4.

**DISCUSSION**

The state of $T > T_s$ in FeSe has been much less revealed than in the nematic phase, partly due to lack of tools appropriate to investigate the nematic fluctuations. In the following, we show that nematic ultrafast dynamics above $T_s$ elucidates surprising details of this largely uncharted territory. As shown in Figs. 1, 2 and 3, when $T$ increases beyond $T_s$, the nematic signatures persist until at least 200 K. In comparison, the magnitude of $\Delta R/R$ for $T > T_s$ is smaller than that for $T < T_s$, but two features of the fast component directly appear. (i) The polarity of $A_1$ reverses sign immediately for $T > T_s$, as seen in Fig. 3a and more clearly with $|A_{1,0}|-|A_{1,90}|$ in Fig. 3g. (ii) With decreasing $T$, the nematic signature in $A_1$ emerges at 200 K, shown in the inset of Fig. 3a. At low temperatures below $T_s$, the FS elongation at the $\Gamma$ point was reported to be with an orientation by 90° relative to the FS elongation direction at the M point (see Fig. 4) [22]. Enlightened by this drastic FS reconstruction, we propose a scenario of the state in FeSe for $T > T_s$ to reconcile both features (i) and (ii). In the high-temperature tetragonal phase, FS of FeSe has $C_4$ symmetry. When temperature is decreased to ~ 200 K, with the FS at the M point retaining $C_4$ symmetry, the nematic fluctuations at the $\Gamma$ point emerges. As FS reconstruction or any fluctuation at the M point is still absent in this temperature range, the QP relaxation changes probed above $T_s$ are dominated by the FS fluctuations at the $\Gamma$ point. The consequent nematic sign of $|A_{1,0}|-|A_{1,90}|$ between 90 K and 200 K is opposite to that below $T_s$, since the orientations of FS elongation at the M and $\Gamma$ point are just opposite below $T_s$. This scenario is shown schematically in Fig. 4. It is noted that at $T < T_s$ the electronic nematicity is dominated by the FS reconstruction at the M point, for the FS reconstruction at the $\Gamma$ point is less severe than at the M point[10,14-16,26-28]. It is also worth noting that, with almost equal $\tau_{1,0}$ and $\tau_{1,90}$ above $T_s$ shown in Fig. 3c, the QP relaxation dynamics along $k_x$ and $k_y$ are almost the same with nematic fluctuations at the Γ point. This effect is in contrast to the case of anisotropic $\tau$ below $T_s$, which is dominated by the FS reconstruction at the M point.

The slow component with spin coupling further unveils previously elusive magnetic properties of the regime for $T > T_s$. At high temperatures, $A_{2,0} - A_{2,90} = 0$ as seen clearly in Fig. 3h. The lack of nematic fluctuations in the slow component indicates that the average magnetic moments along $x$ and $y$ directions are the same, reflecting the continuous rotational symmetry. Nevertheless, with decreasing $T$, the nematic signal of $A_{2,0} - A_{2,90} \neq 0$ clearly shows at about $T = 150$ K (see Fig. 3h and the inset). This nematic signal of $A_2$ indicates the onset of a magnetic subsystem breaking the four-fold symmetry. Very recently, the existence of magnetic fluctuations at $T = 110$ K has been reported by inelastic neutron scattering [46]. However, the existence of magnetic fluctuations between $T_s$ and up to at least 150 K has not been discovered until in the present work. Moreover, the sign of $A_{2,0} - A_{2,90} < 0$ is opposite to that below $T_s$, indicating that the fluctuating spins tend to align along the $y$ direction. This rotation of the spin direction by 90° above $T_s$ is coupled to the fluctuating FS elongation at the Γ point with an orientation by 90° relative to that at the M point below $T_s$. (An alternative origin of these differences is the sign inversion of the orbital polarization at the Γ point unique to FeSe, which produces $n_{xz} > n_{yz}$, in contrast to the FS reconstruction at the M point below $T_s$. The orbital polarization at the $\Gamma$ point is related to the band splitting which existence above $T_s$ was observed by ARPES[28]. The origin of the band splitting at the $\Gamma$ point remains elusive[28], but is unlikely due to nematic fluctuations for $T > T_s$ since regular ARPES probes static orders.) Overall, we have observed both nematic and magnetic fluctuations in FeSe at high temperatures. The coupling between the magnetic fluctuations and the change of the electronic structure at high temperatures is also identified, as below $T_s$. The regime with the newly discovered nematic and magnetic fluctuations is denoted by the yellow area in Fig. 4. Careful measurements of the magnetic properties support an onset of the magnetic fluctuations at a temperature far above $T_s$, as shown in S4. The distinct FS elongation directions at the M and $\Gamma$ points weaken the FS nesting. This effect is likely a key to the absence of static magnetic order in FeSe. There are more discussions in S5 on the temperature range of the nematic and magnetic fluctuations above $T_s$. The onset temperature of the magnetic fluctuations is, notably, near $T^*$, at which the slope of $\rho(T)$ demonstrates a rapid change (see Fig. 4). These results hint at a nematic/magnetic origin of $T^*$.

Finally, the nematic ultrafast QP dynamics in FeSe has been thoroughly studied by polarized pump-probe spectroscopy. Two distinct relaxation components were observed in $\Delta R/R$. The fast component on the time scale 0.1-1.5 ps is associated with the electronic structure and the slow component on the time scale 8-25 ps is assigned to the energy relaxation through the spin channel together elucidate an exotic phase diagram of FeSe shown in Fig. 4, where both nematic fluctuations and an elusive spin subsystem are hidden above $T_s$. The present results certainly inspire a possible scenario for all Fe-based superconductors, which needs to be confirmed by other probes.

**METHODS**

We grew FeSe single crystals in evacuated quartz with a KCl-AlCl$_3$ flux technique [51]. The crystalline structure and transport properties of the samples were examined by x-ray diffraction and van der Pauw measurements, respectively. The femtosecond spectroscopy measurement was performed with a dual-color pump-probe system (for the femtosecond laser, the repetition rate 5.2 MHz, wavelength 800 nm, pulse duration 100 fs) and an avalanche photodetector with the standard lock-in technique. This non-degenerate pump-probe scheme can significantly eliminate the annoying coherent spike around zero time delay [52]. The fluences of the pump and probe beams were 39.7 and 2.3 µJ/cm$^2$, respectively. The pump pulses have a corresponding photon energy (3.1 eV) at which greater absorption occurred in the absorption spectrum of FeSe [53], and hence generate electronic excitation. To study the QP dynamics we measured the photoinduced reflectivity ($\Delta R/R$) transients of the probe beam with photon energy 1.55 eV. $\Delta R/R(t, \phi_{\text{pump, probe}})$ curves along various orientations on the surface of the sample were obtained on rotating the polarization of pulses at nearly normal incidence ($\theta_{\text{pump}} \sim 0°$, $\theta_{\text{probe}} \sim 7°$). The intensity and polarization (electric field, **E**) of pulses were adjusted with a λ/2 plate and polarizer [38-41]. Moreover, the penetration depth of FeSe is ~ 24 nm for 400 nm and ~30 nm for 800 nm, which are estimated from the skin depth of electromagnetic wave in metal, $\lambda/4\pi k$ [11].

The spot size of the probe beam in this study is 83 μm × 45 μm, which is smaller than the typical domain size of ~ 400 μm × 200 μm in our FeSe single crystals, as shown in Fig. S1 of Supplementary Information. Due to the external stress, e.g., caused by the glue/holder, each domain has its own preferred orientation once the nematicity and even nematic fluctuations appear. Additionally, while the general static measurement (e.g., ARPES) is a probe into the static orders of FeSe, the transient pump-probe spectroscopy is capable of probing the orders that fluctuate fast or the short-range order. This is because, after the fluctuation order is destroyed by a pumping pulse, we can immediately probe the reforming fluctuation order within femtosecond timescale. However, the general static measurements do not provide enough time-resolution to resolve these fast fluctuations and only can obtain the long-time average results, which is usually zero. Moreover, there is no trigger signal (served by a pump pulse) to be a reference point in time domain for the general static measurements.

**DATA AVAILABILITY**

The authors declare that the data supporting the findings of this study are available within the paper and its Supplementary Information files.

**ACKNOWLEDGEMENTS**

This work is supported by the Ministry of Science and Technology of the Republic of China, Taiwan (Grant No. 102-2112-M-009-006-MY3, 103-2923-M-009-001-MY3, 103-2628-M-00-002-MY3, 103-2119-M-009-004-MY3, 103-2119-M-009-007-MY3, 103-2112-M009-015-MY3) and the Grant MOE ATU Program at NCTU. This work was also supported in part by RFBR Grant 14-02-92002 and from the Ministry of Education and Science of the Russian Federation in the framework of Increase Competitiveness Program of NUST «MISiS» (№ K2-2015-075 and K4-2015-020) and by Act 211 Government of the Russian Federation, contract № 02.A03.21.0006. We gratefully acknowledge the assistance of C.-M. Cheng (NSRRC) for the manuscript.

**COMPETING INTERESTS**

The authors declare no competing financial interests.

**CONTRIBUTIONS**

C.W.L. developed the polarized pump-probe spectroscopy. P.C.C. and S.H.W. performed the pump-probe experiments and analyzed the data. J.C.C. carried out the $\chi(T)$ measurements of the samples. D.A.C., O.S.V., and A.N.V. synthesized and characterized the FeSe crystals. The manuscript was written by C.W.L. and J.Y.L. with

**FIGURE LEGENDS**

**Fig. 1** 3D plot of orientation-dependent photoinduced reflectivity ($\Delta R/R$) transients at various temperatures. Inset: schematics of the experimental setup of polarization-dependent pump-probe spectroscopy. $\phi_{pump} = 0°$ and $\phi_{probe} = 0°$ indicate that the **E** field of pump and probe pulses along *a*-axis of an FeSe single crystal.

**Fig. 2 a-d** $\Delta R/R$ of an FeSe single crystal with the polarizations of pump and probe beams along $\phi = 0°$ and 90° at various temperatures.

**Fig. 3** Temperature dependence of the amplitudes **a** $A_1$, **b** $A_2$ and the relaxation times **c** $\tau_1$, **d** $\tau_2$ of $\Delta R/R$ along $\phi=0°$ and 90° resulting from the fits by Eq. (1). Solid lines are fits to the Rothwarf-Taylor model in (**a**), (**b**), and (**c**) (see section S3 and S4 in Supplemental Information). Inset of (**a**) shows the temperature-dependent $A_1$ on an enlarged scale. Inset of (**d**) shows the $\Delta R/R$ along $\phi=0°$ and 90° at 30 K as an example, fit by Eq. (1). **e** Amplitude $A_1$ of (**a**) below $T_s$ fitted with the Rothwarf-Taylor model (solid line). The solid starts show the temperature-dependent band splitting along $k_y$ at M point obtained by ARPES [27]. **f** The band structure along $k_x$ and $k_y$ at M point for $T<T_s$ (thick-solid lines) and $T>T_s$ (thin-dashed lines) [27]. **g** Difference between $A_{1,0}$ ($\phi=0°$) and $A_{1,90}$ ($\phi=90°$) in (**a**). Inset shows the temperature range above $T_s$. **h** The difference

between $A_{2,0}$ ($\phi=0°$) and $A_{2,90}$ ($\phi=90°$) in (**b**). Inset shows the enlarged part above $T_s$. The error bars are the standard deviations estimated from several measurements.

**Fig. 4** Phase diagram of FeSe by nematic ultrafast dynamics. Temperature dependence of the resistivity $\rho$ shows clearly an anomaly at $T_s$ and indicates the high quality of FeSe together with a large residual-resistance ratio (RRR). $T^*$ denotes the temperature at which $\rho(T)$ shows a rapid change of slope. Insets illustrate the nematic evolution of charge and spin subsystems in various phases. The thin arrows indicate sketchily the individual moment of Fe ions. The thick arrows indicate the "net" magnetic moments of FeSe in the stripe form. The simplified FS in each temperature is depicted. The picture of FS for $T<T_s$ follows Ref. 49. The dashed green line denoted the proposed FS fluctuations at the Γ point.

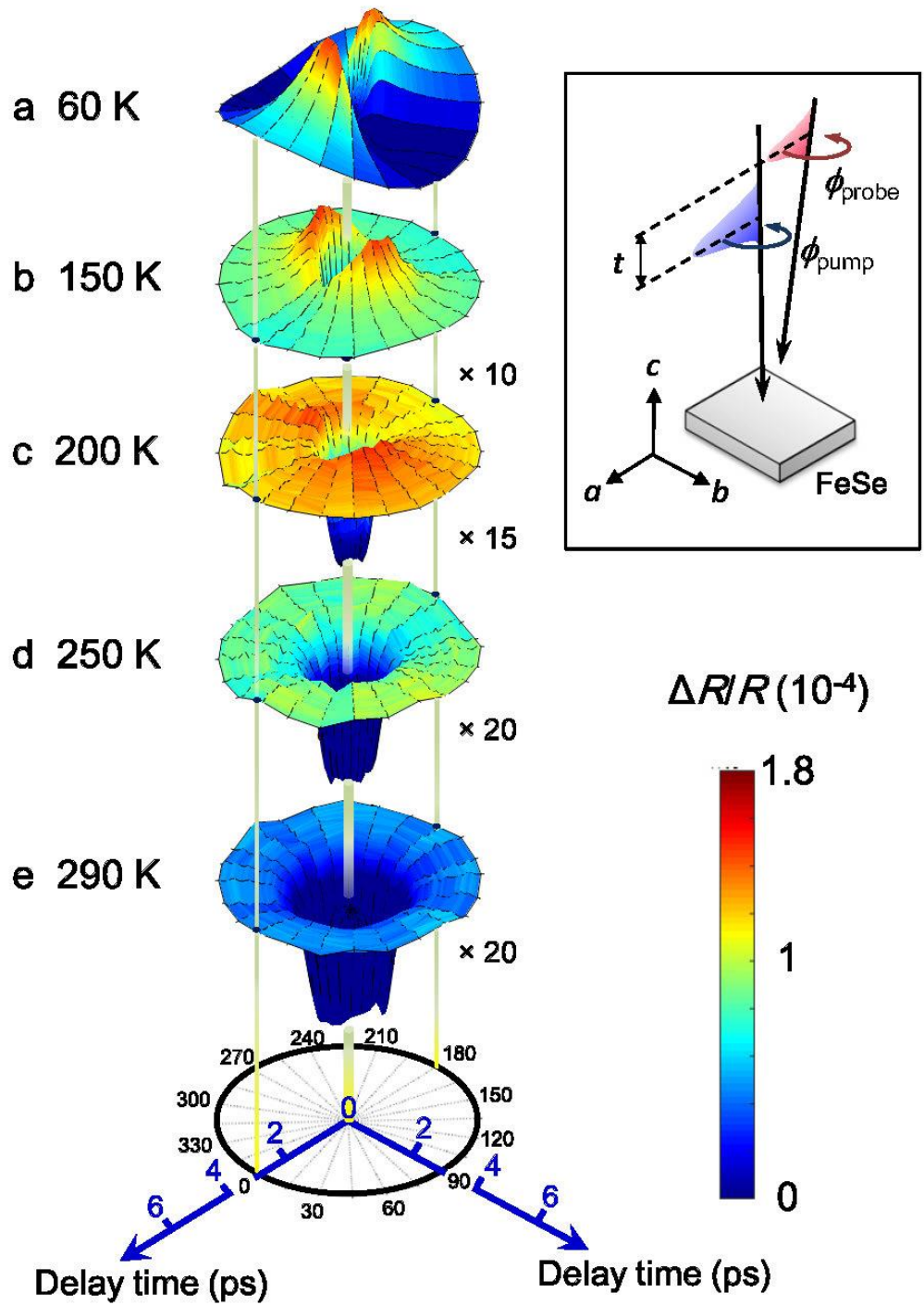

Fig. 1

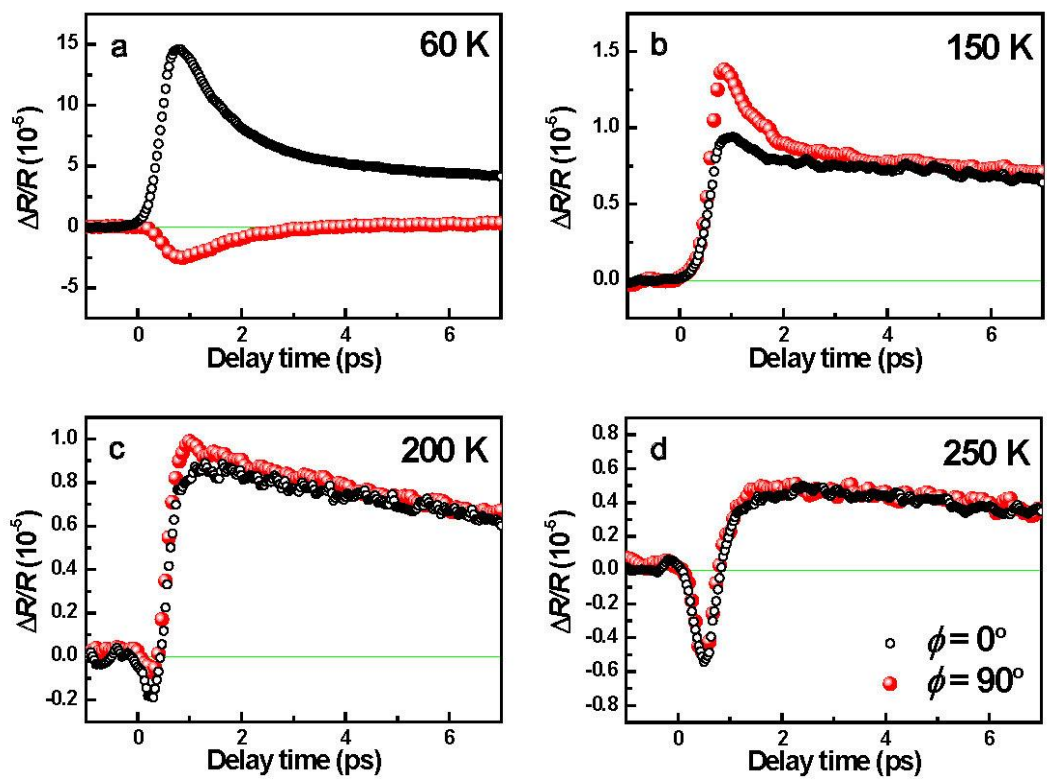

Fig. 2

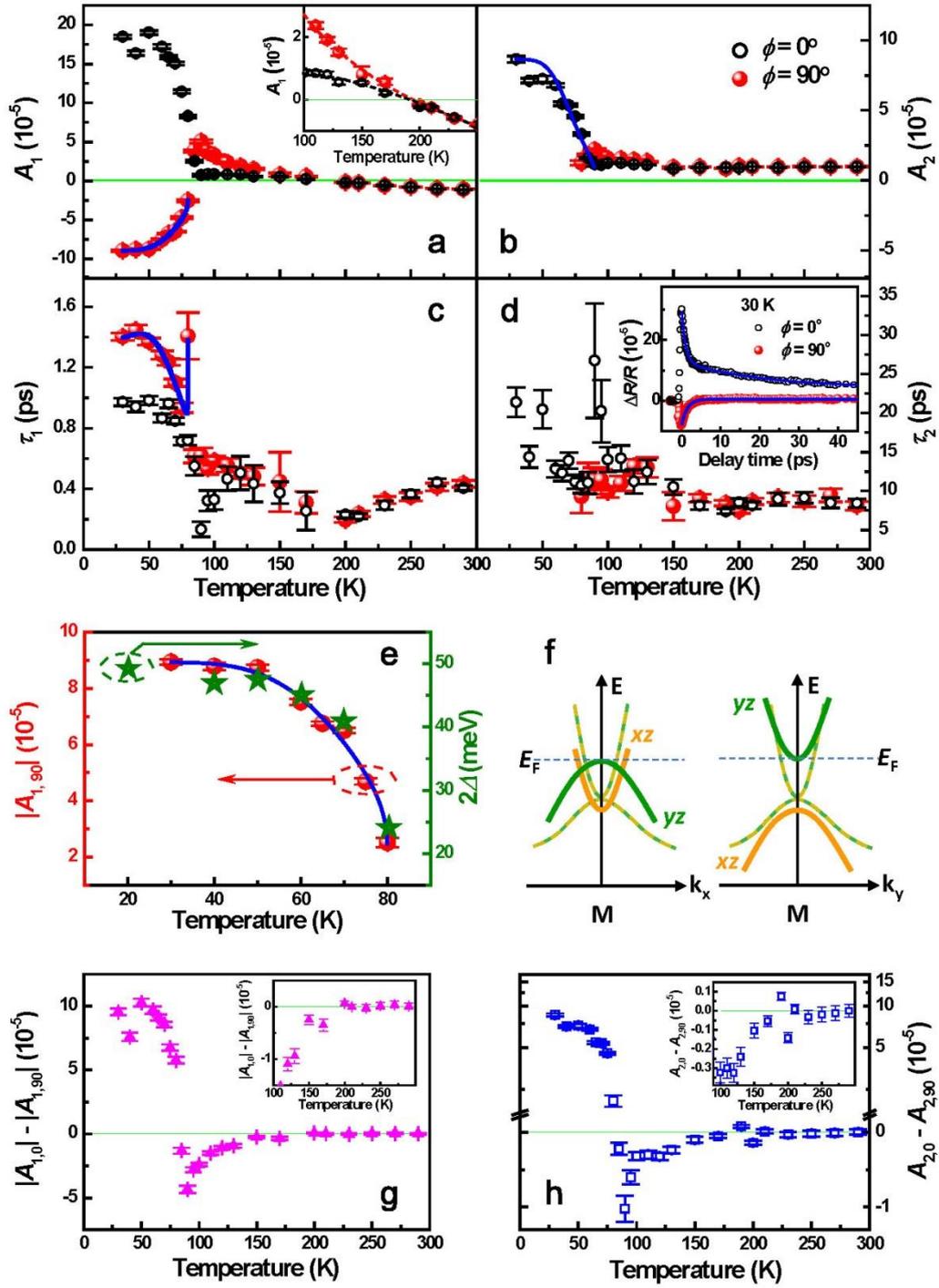

Fig. 3

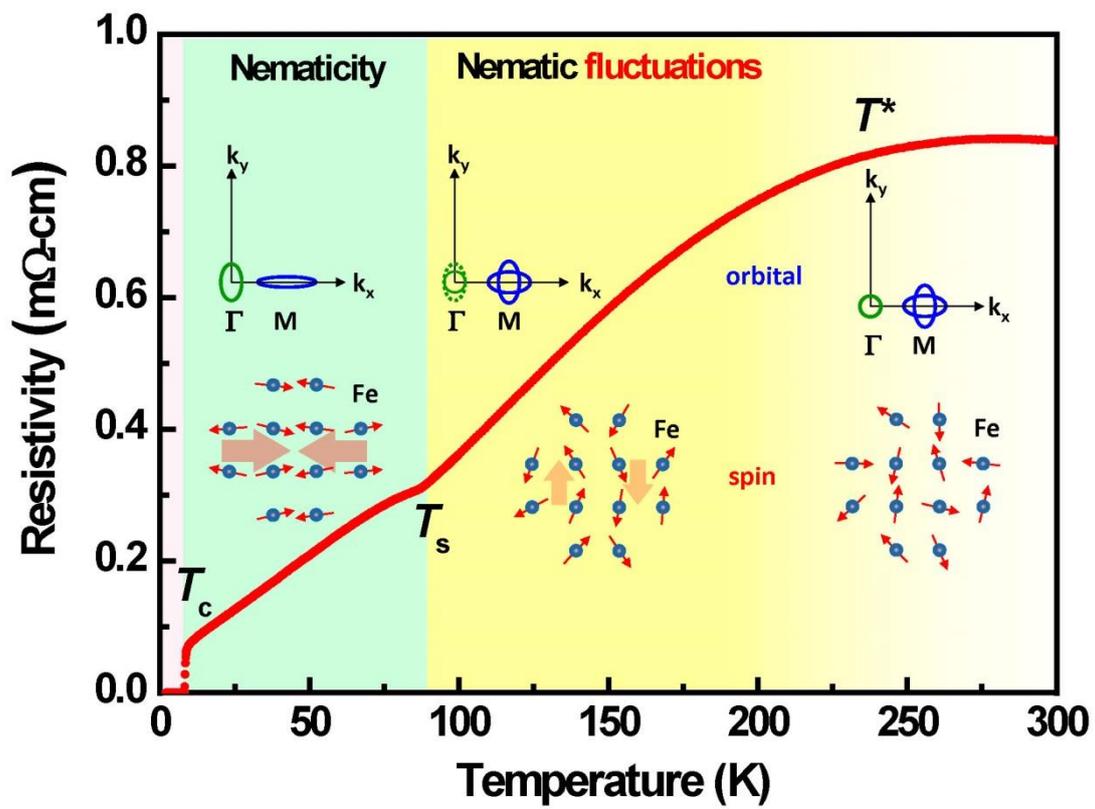

Fig. 4

# Unveiling the hidden nematicity and spin subsystem in FeSe

Chih-Wei Luo, Po Chung Cheng, Shun-Hung Wang, Jen-Che Chiang, Jiunn-Yuan Lin, Kaung-Hsiung Wu, Jenh-Yih Juang, Dmitry A. Chareev, Olga S. Volkova and Alexander N. Vasiliev

**Supplementary Information**

## S1. Spatial mapping of photoinduced reflectivity (Δ*R*/*R*) transients

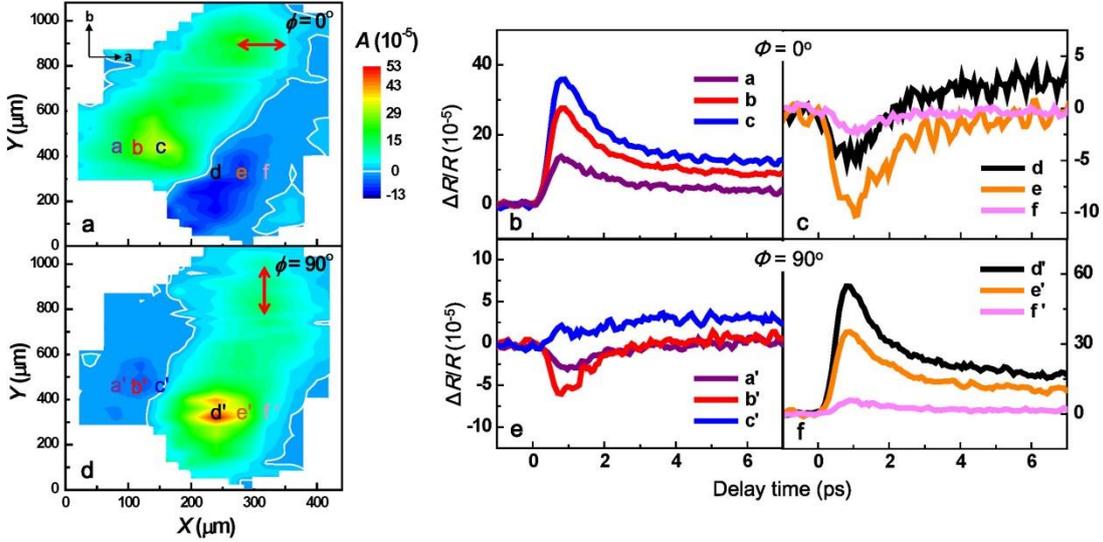

**Figure S1.** Spatial mapping of the peak amplitude of photoinduced reflectivity (Δ*R*/*R*) transients with polarizations of pump and probe beams along (**a**) $\phi = 0°$ (represented by an arrow) and (**d**), $\phi = 90°$ (represented with an arrow) at 30 K. The spot size of the pump beam is 96 μm × 87 μm. The spot size of the probe beam is 83 μm × 45 μm. Fig. (**b**) and (**c**) represent the Δ*R*/*R* transients at points a, b, c, d, e, f in (**a**). Figs. (**e**) and (**f**) represent the Δ*R*/*R* transients at points a', b', c', d', e', f' in (**d**).

## S2. Ascriptions of the first and second terms in Eq. (1)

In FeSe, the electronic excitations generated by the pump pulses result in a rapid rise of Δ*R*/*R* at zero time delay, as shown in Fig. 2. The observed excitation is triggered by transferring the electrons from *d* valence band of Fe to *d* conduction band of Fe [1]. At zero time delay, the number of the excited electrons generated in this non-thermal process is related to the amplitude of Δ*R*/*R*. These high-energy electrons accumulated in the *d* conduction band of Fe release their energy through coupling with the longitudinal-optical (LO) phonons within a couple of picoseconds, which is expressed by the first term in Eq. (1) and the blue dashed lines in Fig. S2. Meanwhile, the hot iterant carriers (or QPs) in metallic FeSe would also transfer their energy to the spin subsystem and then disturb the spin ordering on the timescale of sub-picosecond. Subsequently, the disordered spins would reorder on the timescale of tens of picoseconds [2,3], which is expressed by the $A_2$ component with two-fold symmetry below $T^*$ in Fig. 4. This spin-related process is described with the second term in Eq. (1) and the green dashed lines in Fig. S2(a)-(f).

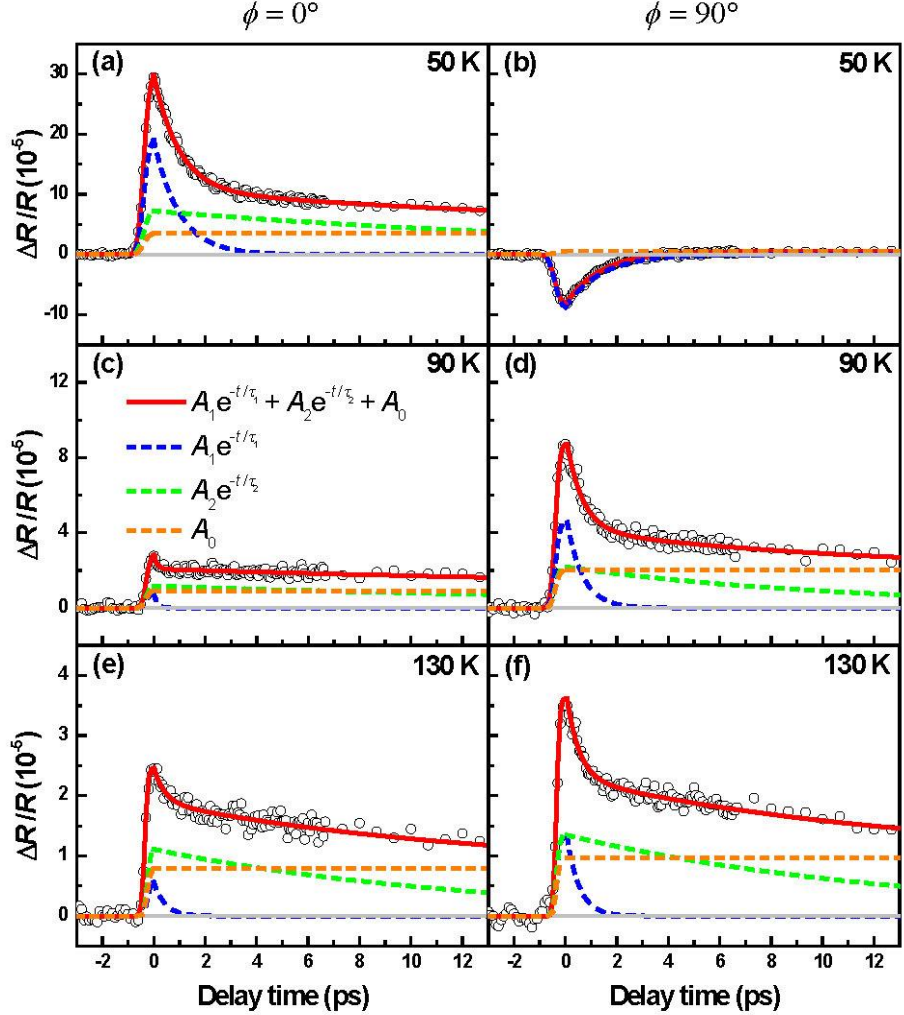

**Figure S2.** The photoinduced reflectivity ($\Delta R/R$) transients along $\phi = 0°$ and $\phi = 90°$ at (a)-(b) 50 K, (c)-(d) 90 K ($T_s$), and (e)-(f) 130 K fitted by Eq. (1). The dashed lines with various colors represent different components in the right-hand side of Eq. (1), respectively. Solid lines represent the sum of all dashed lines. The blue dashed lines: the first term in the right side of Eq. (1). The green dashed lines: the second term in the right side of Eq. (1). The orange dashed lines: the last term in the right side of Eq. (1).

As shown in Fig. S2(e), the $\Delta R/R$ transient with $\phi = 0°$ at 130 K possess one fast relaxation channel (blue dashed line) and one slow relaxation channel (green dashed line). Similarly, the $\Delta R/R$ transient with $\phi = 90°$ (Fig. S2(f)) also show two relaxation channels as we observed at $\phi = 0°$. The difference between these two $\Delta R/R$ transients at $\phi = 90°$ and $0°$ is only the amplitude, which is caused by the Fermi surface (FS) distortion at Γ point as shown in the inset of Fig. 4. Above argument is also applied to both fast and slow relaxation channels at $T > T_s$ and even $T = T_s$ (= 90 K). However, when temperature is lower than $T_s$ (= 90 K), the amplitude of fast relaxation channel at $\phi = 0°$ is larger than that at $\phi = 90°$, which is opposite to the cases at $T \geq T_s$. This is because that the electronic relaxation (fast relaxation channel) is dominated by the FS distortion at M point, whose distortion is much serious than that at Γ point and rotated by 90°, as shown in the inset of Fig. 4.

Additionally, when $T < T_s$, the slow relaxation channel, which pertains to spin reordering through the spin-phonon coupling, totally disappears in the $\Delta R/R$ transient at $\phi = 90°$. This means that the relaxation channel through spin-phonon coupling is broken along $\phi = 90°$, but it is still surviving along $\phi = 0°$. If the phonon is isotropy, this huge anisotropy observed in the slow relaxation channel at $T < T_s$ should come from the spin subsystem, which is not the necessary consequences of FS distortion.

For the present case of FeSe, when the two-fold symmetry in $A_2$ component was established at $T_s$, the relaxation time $\tau_2$ of $A_2$ component simultaneously shows a dramatic increase around $T_s$, which is consistent with the observations in Co-doped BaFe$_2$As$_2$ [4] and Sm(Fe,Co)AsO [5], indicating that the $A_2$ component of $\Delta R/R$ in FeSe is indeed associated with the spin subsystem. To summarize, the slow component $A_2$ in $\Delta R/R$ of FeSe has been experimentally observed in pnictides in the literature, albeit with limited discussions. Intriguingly, the comparisons with the slow component in the relevant literature of pnictides further support the assignment of the slow component $A_2$ to the magnetic subsystem in FeSe.

## S3. Electronic energy gap according to the Rothwarf-Taylor model

After a pump pulse, the electrons are excited from valence band to conduction band at Γ point (i.e., zero wave-vector point). Then, these photoexcited quasiparticles (QPs) will relax at Γ point or scatter to M point through intervalley scattering. Consequently, the relaxation of QPs at M point will suffer the bottleneck effect due to a gap opening. The number of thermally excited QP $n_T(T) \propto [A_{1,90}(T\rightarrow 0)/A_{1,90}(T)]-1$. The temperature-dependent behavior of $n_T(T)$ is further fitted by $n_T(T) \propto [\Delta(T)T]^{1/2}\exp[-\Delta(T)/T]$ where $\Delta(T)$ is the energy band gap. Assuming a mean-field-like temperature-dependent $\Delta(T) = \Delta(0)[1-(T/T_s)]^x$, the fits lead to $2\Delta(0) = 8.14k_BT_s = 56$ meV and $x = 0.178$; $T_s$ is fixed at 80 K. This choice of $T_s$ is to avoid the large background contribution to the fitting from the positive fast component $A_{1,90}$ at temperatures higher than 80 K. It should be noted that $T_s$ in the above fitting would reflect the band gap opening at the M point due to band splitting; therefore it might be slightly lower than the real structure transition temperature at which FS reconstruction just begins to emerge. All of above parameters were also applied to the fitting of $\tau_{1,90}$ in Fig. 3c. The fitting result is in accord with the energy splitting between $d_{yz}$ and $d_{xz}$ near the M point in the Brillouin zone revealed from ARPES [6,7]. Very recently, this gap was assigned to the splitting ($\Delta_M$) between $d_{xz/yz}$ and $d_{xy}$ bands at the M point [8,9]. However, the gap amplitude $\Delta_M$ between $d_{xz/yz}$ and $d_{xy}$ bands, and its temperature evolution are also consistent with temperature-dependent $|A_{1,90}|$ in Fig. 3e and previous ARPES results. The presence of a gap in the QP density of states gives rise to a bottleneck for carrier relaxation, which is clearly observed in the relaxation time $\tau_{1,90}$ near $T_s$. The mechanism of the bottleneck is also describable with the Rothwarf-Taylor model [10]. When one QP with energy higher than $2\Delta$, a high energy boson (HEB) with energy $\omega \geq 2\Delta$ is created. The HEBs that remain in the excitation volume subsequently excite additional carriers below the band gap, effectively preventing QPs from recombination. Until $\omega < 2\Delta$ and the carriers below the band gap are not excited further by HEB, the number of QP finally decreases in several picoseconds. In the case of a mean-field-like gap, i.e., the gap gradually shrinks with $T$ approaching $T_s$ from the low-temperature side, which is consistent with the ARPES results as shown by the solid start symbols in Fig. 3e [6]; more HEB are available to regenerate QPs. The relaxation processes become less and less efficient. Hence, $\tau_{1,90}$ below $T_s$ exceeds 1 ps and diverges around $T_s$, which is dominated by the band splitting along $k_y$ at M point related to the structural phase transition [7].

## S4. The energy scale of magnetic fluctuations

In order to estimate the energy scale of magnetic fluctuations, we fitted the temperature-dependent $A_2$ by the Rothwarf–Taylor model (for the details, please see section S3), which has been used for the temperature-dependent $A_{1,90}$ in Fig. 3a. For the temperature-dependent $A_{2,0}$ in Fig. 3b, the fitting with Rothwarf-Taylor model leads to an energy scale of $2\Delta_m(0) = 72$ meV and $x = 0.216$; The characteristic temperature $T_m$ is fixed at 90 K as required by the experimental data points. The energy scale of 72 meV from the fitting is consistent with the value obtained from inelastic neutron scattering [11]. $T_m$ is associated with the spin subsystem and is not necessary identical to $T_s$ in **S3**.

## S5. Onset temperatures of nematic and magnetic fluctuations above $T_s$

Figure S3a shows the temperature-dependent ($|A_{1,0}|-|A_{1,90}|$) and ($A_{2,0}-A_{2,90}$) above $T_s$. The onset temperature of non-zero ($|A_{1,0}|-|A_{1,90}|$), which is associated with the charge subsystem, is 200 K. The temperature-dependent ($A_{2,0}-A_{2,90}$) associated with the spin subsystem in the high-temperature region reveals the onset (marked by an arrow in Fig. S3a) about 230 K, which is completely consistent with the onset temperature of slope change in the temperature-dependent magnetic moment of Fig. S3b. This feature at 230 K was revealed in our previous work by non-polarization-resolved pump-probe spectroscopy [12], which further infers the opening of a spin gap in FeSe (see S4). Moreover, when the temperature is 200 K, the nematicity of the charge subsystem becomes established (because of $|A_{1,0}|-|A_{1,90}| \neq 0$) and simultaneously couples to the spin subsystem to cause significant fluctuations in the ($A_{2,0}-A_{2,90}$) signal.

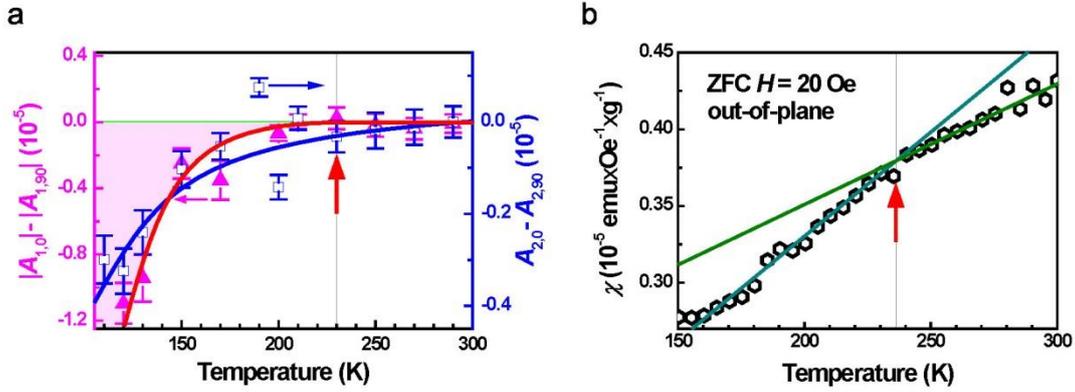

**Figure S3.** (**a**) Difference between amplitude $A_{1,0}$ ($A_{2,0}$) and $A_{1,90}$ ($A_{2,90}$) in Fig. 3a (Fig. 3b) above $T_s$. The magenta area represents the onset temperature of ($|A_{1,0}|-|A_{1,90}|$). (**b**) Temperature-dependent magnetic susceptibility in an FeSe single crystal above $T_s$. Solid lines are guides to the eyes. The arrows indicate the onset temperature of non-zero ($A_{2,0}-A_{2,90}$) and the slope change in $\chi(T)$.

## S6. Orientation dependence of the relaxation times $\tau_1$ and $\tau_2$ at various temperatures

To reveal the overall anisotropic dynamics of electronic structure and the spin subsystem in FeSe, we plot the orientation-dependent relaxation times extracted from Fig. 1 in Figs. S4a and S4b. The relaxation of QPs through coupling with phonons shows a strong anisotropy at 60 K on (001) plane of FeSe. The significantly enhanced $\tau_1$ is caused by the gap opening along $k_y$ at the M point as illustrated in Fig. 3f [6]. In contrast, the relaxation of excitation energy through coupling with spins is also strongly orientation-dependent at 60 K, indicating that the spins mainly align along $k_x$ (see the inset of Fig. 4) [13]. While the temperature is above $T_s$, this remarkable anisotropy of relaxation time drastically shrinks. For $\tau_2$, its anisotropy persists to temperatures between 200 K and 250 K to indicate that the spin nematicity in FeSe can be observed above 200 K, i.e., the temperature marked by the arrows in Fig. S3. The anisotropy of $\tau_1$ is observable only until 200 K. This comparison indicates that the nematicity of spin subsystem appears at temperatures higher than that of the electronic nematicity concurring with Fig. S3a.

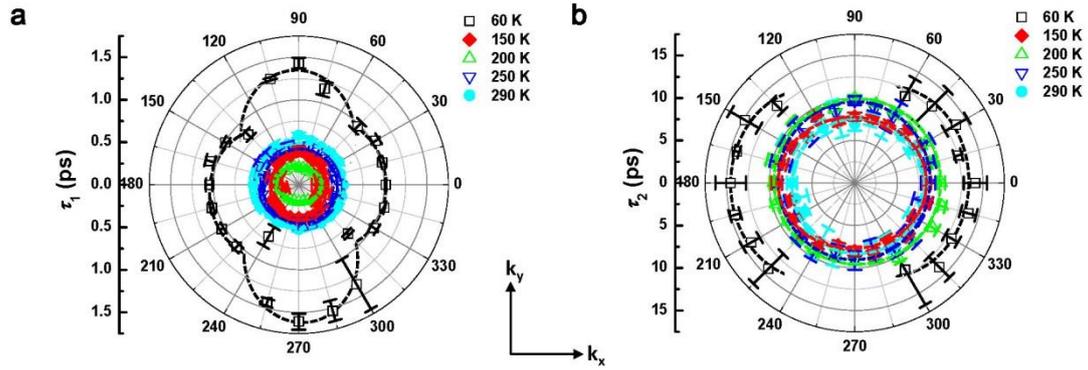

**Figure S4.** Orientation dependence of relaxation time (**a**) $\tau_1$, (**b**) $\tau_2$ at various temperatures from fitting Eq. (1) in Fig. 1. Dashed lines are guides to the eyes.

## S7. Pumping fluence dependence of the photoinduced reflectivity ($\Delta R/R$) transients

As shown in Fig. S5, the amplitude of $\Delta R/R$ transients linearly rises as increasing the pumping fluence. By normalizing, all of the $\Delta R/R$ transients overlap together to show the same relaxation behavior (see the inset of Fig. S5). Therefore, the $\Delta R/R$ transients of FeSe are pumping fluence-independent while the pumping fluence is below 64.1 $\mu J/cm^2$.

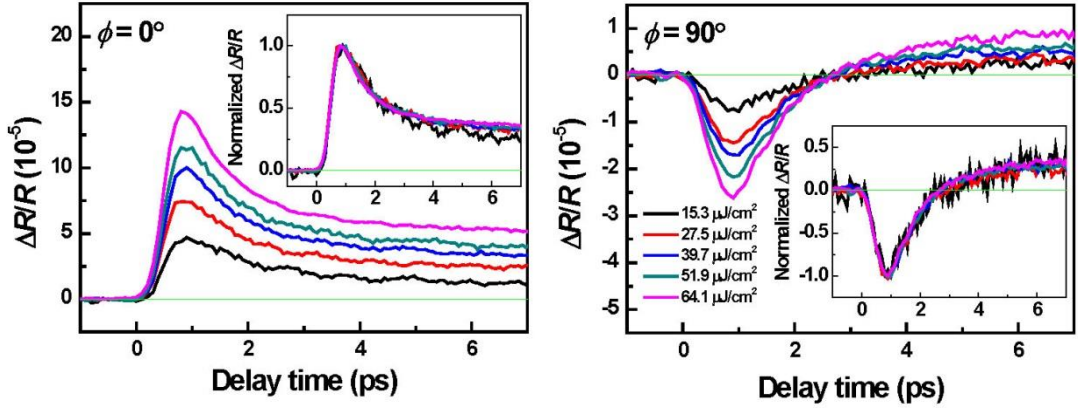

**Figure S5.** Pumping fluence dependence of Δ*R*/*R* transients at 60 K. Inset: the normalized Δ*R*/*R* transients.

## Supplement References